# Why does maximum $T_c$ occur at the cross-over from weak to strong electron-phonon coupling in high temperature superconductors?

Dragan Mihailovic

*Jozef Stefan Institute, Jamova 39, Ljubljana, SI-1000 Slovenia*

*Abstract*

In cuprate superconductors, a pronounced maximum of superconducting $T_c^{max}$ is observed in compounds that have an in-plane Cu-O distance $a_{Cu-O}$ close to ~1.92 Angstroms. On the other hand, direct measurements of the electron-phonon coupling $\lambda\langle\omega^2\rangle$ as a function of $a_{Cu-O}$ show a clear linear correlation, implying that $T_c^{max}$ is a strongly non-linear function of $\lambda\langle\omega^2\rangle$. Conventional superconductivity theories based on the electron-phonon interaction predict a monotonic dependence of $T_c^{max}$ on $\lambda\langle\omega^2\rangle$, which makes them incompatible with the observed behavior. The observed cross-over behavior as a function of $\lambda\langle\omega^2\rangle$ suggests that $T_c^{max}$ occurs at the cross-over from weak to strong coupling, which is also associated with the onset of carrier localization. A coexistence, with a dynamical exchange of localized and itinerant carriers in a two-component superconductivity scenario are in agreement with the observed anomalous behavior and are suggested to be the key to understanding the mechanism for achieving high $T_c^{max}$.

## *Introduction: lack of progress on understanding $T_c$.*

Superconductivity at *low* temperatures is not a rare phenomenon. Most elements (33 at current count) are superconducting at ambient pressure, and a ~20 more are superconducting under pressure. The discovery of superconductivity in oxides by K. Alex Muller and G. Bednorz kick-started a bonanza searching for new compounds with *high* T$_c$. Although myriads of *similar* compounds showed superconductivity, no significant new idea had emerged on how to synthesize a new superconductor with a high transition temperature beyond those already known before HTS cuprates were discovered. Curiously, at the most recent M²S conference (in 2019), amongst hundreds of presentations, only a handful addressed the most crucial question of how to reach high $T_c$s. The problem almost seems to have been abandoned.

Early measurements of the isotope effect in optimally doped YBa$_2$Cu$_3$O$_7$ and EuBa$_2$Cu$_3$O$_7$ suggested that the lattice was not involved in superconductivity [1], which led to a significant amount of theoretical work on purely electronic mechanisms. But despite very intense efforts in this direction, there is so far no answer to the question: what is the mechanism to reach high critical temperatures? Subsequent systematic measurements of the isotope effect by Hugo Keller's group, amongst others, showed a significant isotope effect [2], implying that the lattice is somehow, unconventionally, involved in superconductivity. Polaron formation and local pairing were thought to play an important role, but detailed theory that can explain the observed phenomenology has not yet reached consensus. More recently, the conventional BCS mechanism of superconductivity was resurrected in superconducting hydrides under pressure, where a more-or-less conventional isotope effect has been observed [3]. But an understanding of the role of the lattice in the mechanism for high $T_c$s is still not reaching substantially beyond Eliashberg theory.

Without a fundamentals-based recipe on synthesizing a new high-temperature superconductor from strongly correlated electronic theories, what remains are older, more conventional strong coupling theories, where the T$_c$ is predicted by McMillan's formula [4], or extended Eliashberg theory [5]. The problem with the former is that the expression for $T_c$ includes a parameter (the pseudopotential $\mu^*$), which is not experimentally measurable, nor is it easy to relate it to any physical property of a material that could usefully serve as a guide in the search for higher critical temperatures. The importance of Eliashberg theory is that it points out the importance of high-frequency phonons contributing to the Eliashberg function $\alpha^2 f(\omega)$, where $\alpha$ is the coupling strength, and $f(\omega)$ is the phonon density of states. This is seemingly important in HTS hydrides that include light atoms in a perovskite-like structure, but does not explain why organic superconductors or intercalated graphite which also have light atoms and a top-heavy $f(\omega)$, have relatively low T$_c$s. Thus, basic Eliashberg theory is seemingly insufficient to systematically explain high T$_c$s in different compounds.

## *Maximum T$_c$s at the crossover from weak to strong coupling*

It has been known for a long time that the maximum $T_c^{max}$ that occurs at optimum doping for different compounds, plotted against Cu-O lattice distance $a_{Cu-O}$ shows a clear maximum at $a_{Cu-O}$ ~1.92 angstroms (Fig. 1a) [6]. It was speculated that possibly this behavior is somehow associated with a peak in the electron-phonon coupling strength, but its origin has not been satisfactorily explained so far. To investigate this idea, we performed systematic measurements of the electron-phonon interaction $\lambda\langle\omega^2\rangle$ in a large number of compounds using femtosecond electron-phonon thermalization measurements [7].

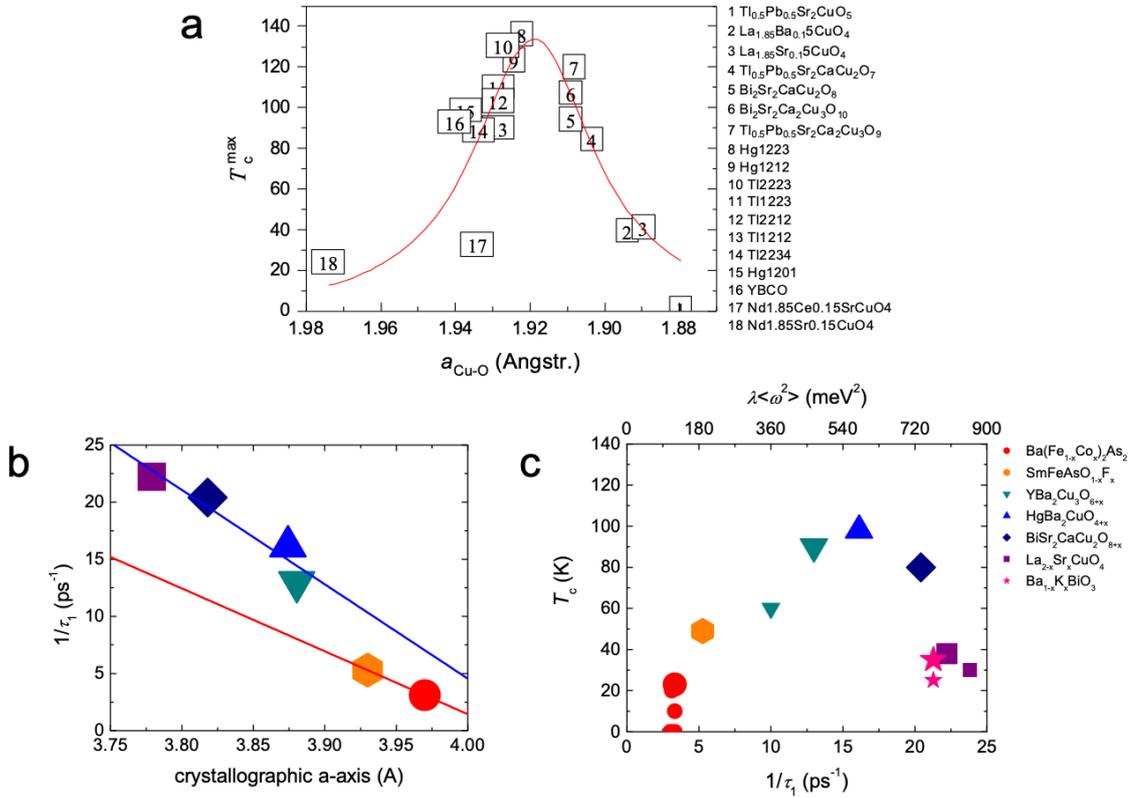

**Figure 1 The evidence for maximum $T_c$s at the crossover from weak to strong coupling:** a) Empirical plot of maximum Tc vs. Cu-O distance $a_{Cu-O}$ in the cuprates. b) electron-phonon (e-p) thermalization rate $1/\tau_1$ vs. $a_{Cu-O}$. The legend for the compounds is shown in c). c) $T_c$ vs. $\lambda<\omega^2>$, and e-p thermalization rate $1/\tau_1$ and Cu-O distance in different HTS superconductors.

Using this technique and the well-tested two-temperature thermalization model [8,9], we are able to establish a linear correlation between the Cu-O distance $a_{Cu-O}$ and $\lambda\langle\omega^2\rangle$ (Fig. 1b). Combining this with the dependence of $T_c^{max}$ on two empirical graphs, we see that the maximum $T_c$ is *not* a monotonically increasing function of $\lambda$. Rather, a plot of $T_c^{max}$ against $\lambda\langle\omega^2\rangle$ shows a clear maximum. The correspondence of the compounds on the plot of $T_c^{max}$ vs. Cu-O distance $a_{Cu-O}$ and $T_c^{max}$ vs. $\lambda\langle\omega^2\rangle$ is worth noting: the same compounds appear in similar positions on either side of the maximum in Figs. 1 a and c. The graph conclusively shows that $T_c^{max}$ does not occur at maximum $\lambda\langle\omega^2\rangle$, but rather, it occurs at the crossover from weak to strong coupling.

*Physics at the cross-over: spatial inhomogeneity and the coexistence of localized and itinerant states*

It was already recognized by Feynman [10] before HTS was discovered that the cross-over from weak to strong coupling is abrupt [11,12]. This means that thermal or quantum lattice fluctuations can easily convert itinerant carriers to localized ones, and vice versa. At finite temperatures, we expect to observe a coexistence of itinerant and localized carriers, which are dynamically resonantly exchanged [11,13]. The experimental observation of coexistence of itinerant and localized states dates back to the early-1990s [14]. Optical spectroscopy revealed a clear and systematic signature of mid-infrared localized polaron spectra in underdoped cuprates [15]. An itinerant-carrier component in the form of a Drude-like response appears upon doping, that forms the condensate below $T_c$, while the polaronic spectral weight remains unchanged. Clear signs of localization were reported from femtosecond pump-probe experiments [16–19]. Numerous other techniques have since revealed the inhomogeneous nature of materials in the normal state in which charge carriers are on the verge of localization, highlighting the complex interplay between localized and itinerant states. Many different competing or intertwined orders (AFM, CDW, pair-density wave etc.) have been reported, but a number of apparent dichotomies also became increasingly apparent, whereby different experiments revealed either a localized-state strongly correlated picture, or seemingly coexisting itinerant carriers within the same material. For example, time-resolved optical measurements that assigned localized carriers to the pseudogap state, coexisting with itinerant states [16–20]. ARPES and STM, which both probe the surface, typically reveal itinerant and localized carriers respectively. ARPES shows dispersive bands, while localized state features in STM are remarkably similar to those predicted by charged lattice gas calculations of polaron self-organization [21,22].

Further support for the coexistence comes from a seemingly universal experimental finding that the SC gap is spatially uncorrelated with inhomogeneity of the pseudogap: the low-energy gap structure associated with SC is much more spatially homogeneous than the higher-energy gap structure associated with the PG temperature scale [23]. The two types of excitations, related to superconducting pairs, and localized particles that are responsible for forming the pseudogap are seemingly spatially uncorrelated.

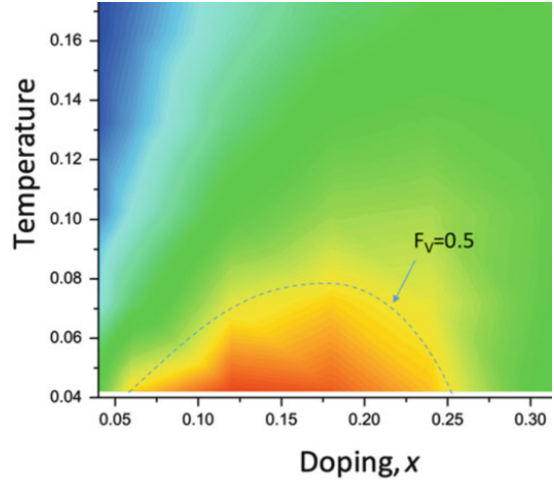

Figure 2 The predicted $T_c$ for a model based on percolation between paired localized states in the 2-component scenario as a function of doping fraction x [22,27]. The color scale represents pair density (red=1, blue=0).

## Achieving high $T_c$ in the mixed state

A simple model for estimating $T_c$ of such a 2-component system, irrespective of the microscopic details of the carrier tunneling between localized inter-site pairs and itinerant carriers, was proposed by Mihailovic, Kabanov and Müller [24]. To achieve macroscopic phase coherence *and* high $T_c$ within the 2-component picture, a finite density of *both* itinerant and localized components needs to be present. Model calculations with a charged lattice gas model (CLG) have shown that for repulsive polarons, a finite inter-site pair density exists over a significant part of the phase diagram, exhibiting a characteristic dome shape as a function of doping [22,25,26] (Fig. 2). The inter-site pair number density is given by the color scale. The $T_c$ (dashed line) corresponds to the percolation threshold reached at a volume fraction $F_V = 0.5$, appropriate for 2D.

The model for calculating $T_c$ considers phase coherence percolation between such local inter-site pairs, predicts the value of $T_c$ in YBa$_2$Cu$_3$O$_{7-x}$ to within <20% of the experimental value, without any adjustable parameters [24]. The actual *value* of $T_c^{max}$ within this model is determined by the intersite pair binding energy, that can be measured by various techniques, and usually referred to as the pseudogap energy, localization energy, or (bi)polaron binding energy. This simple model gives us a general understanding of the behavior of $T_c$ versus doping and the EPI strength $\lambda\langle\omega^2\rangle$ for a system with coexisting itinerant and localized charges on the sole assumption that superconductivity occurs by phase coherence percolation via localized inter-site pairs. Macroscopic phase coherence is assumed to occur via resonant tunneling between the localized pair states and the itinerant carrier reservoir. The details of this process will not have a significant effect on the phase diagram, or the value of $T_c^{max}$ derived on the basis of the percolation threshold.

*Conclusions*

The main message of this paper is to highlight the importance of the coexistence of localized and itinerant carriers that gives rise to the empirically observed, non-trivial, non-monotonic dependence of $T_c$ on $\lambda\langle\omega^2\rangle$ in cuprate HTS. The maximum $T_c$ occurs at the crossover from weak to strong coupling, where exchange between itinerant and localized carriers is caused by fluctuations. Increased localization at higher $\lambda\langle\omega^2\rangle$ decreases the pair density in favor of larger localized clusters, while at smaller $\lambda\langle\omega^2\rangle$, the dominance of itinerant carriers leads to a more conventional superconductivity. Now the outstanding experimental challenge is to record spatially-resolved 'snapshots' of the local lattice deformations and charge dynamics associated with pairing, and the carrier conversion dynamics from itinerant to localized states in a direct way.

*Acknowledgments*

On this special occasion of his 95[th] birthday I wish to acknowledge K. Alex Müller for continuing inspiration, valuable discussions and encouragement over the years that kept me working in this remarkable field for such a long time. I also wish to acknowledge V.V. Kabanov and T. Mertelj for contributions to the presented and cited work.